\documentclass[twoside,fleqn]{ActaStyle}
\usepackage{times,cite}

\usepackage{lscape}             
\usepackage{graphicx,epsfig}    
\usepackage[tbtags]{amsmath}    

\def\authorheading{T.~Sakaguchi et al.}

\def\shorttitle{Photon production at RHIC-PHENIX}

\begin{document}

\pagerange{1}{6}

\title{%
Photon production in $\sqrt{s_{NN}}$=200\,GeV Au+Au collisions measured by the PHENIX experiment at RHIC.
}

\author{
T.~Sakaguchi\email{takao@bnl.gov}, for the PHENIX Collaboration
}
{
Brookhaven National Laboratory, Physics Department, Upton, NY 11973, U.S.A.
}

\day{December 12, 2005}

\abstract{%
Direct photon production in Au+Au collisions at $\sqrt{s_{NN}}$=200\,GeV
has been measured in the $p_T$ range of 1$<$$p_T$$<$5\,GeV/$c$. The yield
as a function of centrality is in agreement with published data of the
RHIC 2002 run, and with results from a new method that explores very low
mass dileptons. The result is compared to several theoretical calculations,
and it is found that the measurement is not inconsistent with calculations
including thermal photon contributions.
}


\section{Introduction}
Several discoveries from the first five years at the Relativistic Heavy Ion
Collider (RHIC) indicate that a new hot and dense partonic matter have most
likely been produced in central Au+Au collisions at
$\sqrt{s_{NN}}$=200\,GeV~\cite{bib1,bib2,bib3,bib4}.
The drastically increased hard scattering
cross-section at the RHIC energy makes it possible to use hard probes (jets,
heavy quarks, etc.) as "measures" of the colliding system. High $p_T$ hadrons,
i.e. fragments of jets, have been found to be suppressed as a consequence of
a final state interaction (energy loss) of hard scattered partons with
the partonic medium. This observation is supported by the high $p_T$ direct
photon measurement by the PHENIX experiment~\cite{bib5}; the high $p_T$
photons originating from an initial hard scattering were not suppressed.
The energy loss phenomena has also been observed with
electrons from semi-leptonic decay of charm and bottom quarks~\cite{bib6}.
In addition to new probes, conventional observables also provided important
information. For instance, a large collective flow of bulk particles gives
a hint of a rapid thermalization of the partonic matter, because a large
asymmetry of pressure gradient that results in a large flow can only be realized
in the early time. In summary, the results suggest that the matter created
in heavy ion collisions at RHIC is strongly interacting and quite opaque like
a perfect liquid.

Provided that the hot and dense partonic matter is created,
its thermodynamical nature such as temperature, phase transition order
parameters and the degree of freedom in the medium are of great interest.
The temperature obtained from a statistical model fit to particle yield
ratios is the chemical freeze-out temperature, and no longer reflecting
the one in the partonic matter. Therefore, more direct probes are deserved.

\begin{figure}[htbp]
\begin{minipage}{63mm}
\centering
\leavevmode\epsfxsize=4.8cm
\rotatebox{-90}{\epsfbox{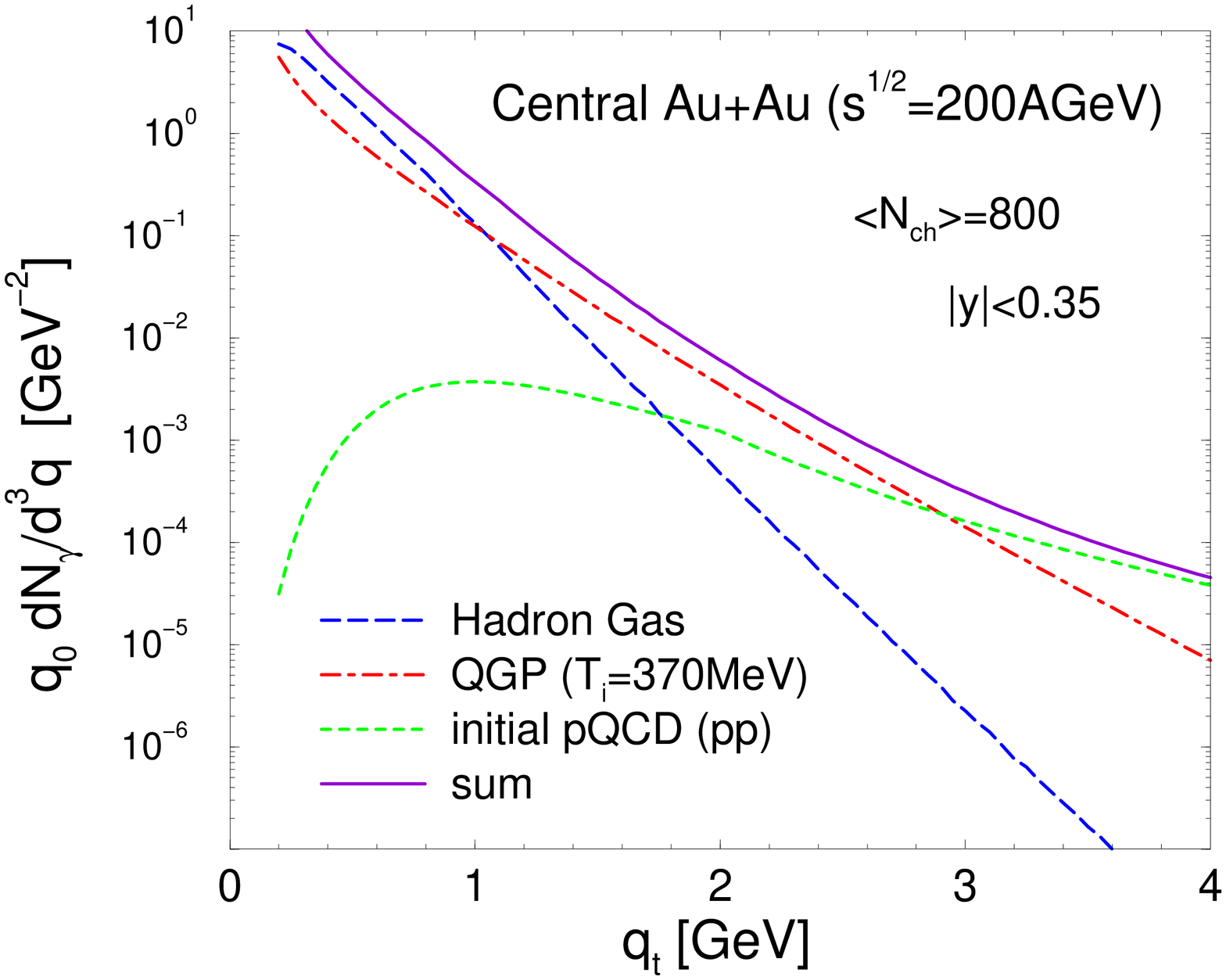}}
\caption{Theoretical expectation on hadron gas interaction, thermal radiation, and pQCD photon for central Au+Au collisions at $\sqrt{s_{NN}}$=200GeV.}
\label{fig1}
\end{minipage}
\hspace{5mm}
\begin{minipage}{63mm}
\centering
\hspace{-7mm}
\leavevmode\epsfxsize=6.5cm
\epsfbox{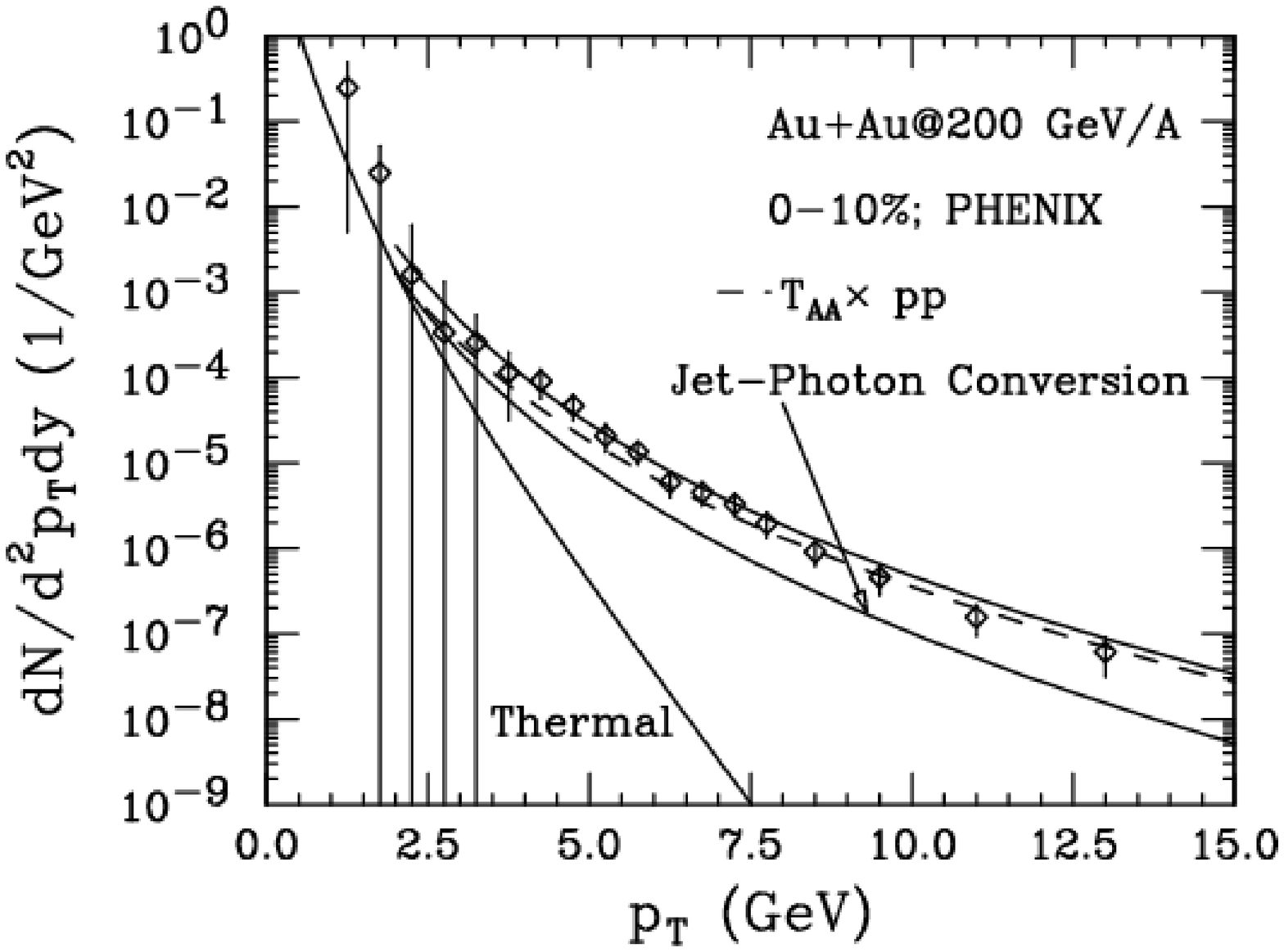}
\caption{Theoretical expectation including jet-photon conversion compared with published data for central Au+Au collisions at $\sqrt{s_{NN}}$=200GeV.}
\label{fig2}
\end{minipage}
\end{figure}

Photons are an excellent probe for extracting such the thermodynamical
information, because once produced, they do not interact strongly with
a medium. They are produced through Compton scattering of quarks and gluons
and annihilation of quarks and anti-quarks (leading order), and bremsstrahlung
or fragment (next leading order). There is also a prediction of a jet-photon
conversion process, which occurs if the hot and dense matter is formed,
by a secondary interaction of a hard
scattered parton with thermal partons in the medium~\cite{bib7}.
The contributions from various processes are mixed up in any $p_T$ range,
but each can dominantly be seen at certain $p_T$'s. The calculation shown
in Fig.~\ref{fig1} predicts that photons with $p_T$$<$1\,GeV/$c$ are mostly
contributed by hadron gas interaction via processes of:
$\pi\pi(\rho) \rightarrow \gamma \rho(\pi)$, 
$\pi K^* \rightarrow  K \gamma$ and etc..~\cite{bib7}
The thermal radiation from QGP state is predominant in 1$<$$p_T$$<$2.5\,GeV/$c$.
For $p_T$$>$2.5\,GeV/$c$, a contribution from jet-photon conversion process
will be seen on top of the one from the initial hard scattering as shown
in Fig.~\ref{fig2}~\cite{bib8}.

The PHENIX experiment at RHIC has succeeded to scope 1$<$$p_T$$<$5\,GeV/$c$
in Au+Au collisions at $\sqrt{s_{NN}}$=200\,GeV where the thermal radiation
and the jet-photon conversion process are likely to manifest. This paper
will present a detailed description of the analysis and the results, together
with theoretical models.

\section{Detector and Analysis}
\begin{figure}[htbp]
\centering
\leavevmode\epsfysize=4.5cm
\epsfbox{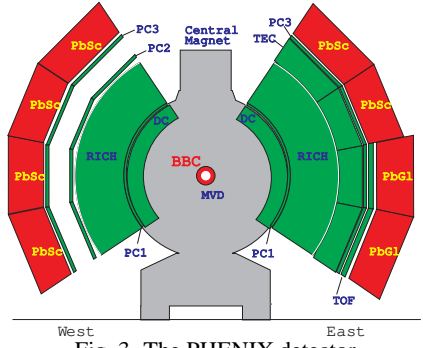}
\vspace{-5mm}
\caption{The PHENIX detector.}
\label{fig3}
\vspace{-5mm}
\end{figure}

Fig.~\ref{fig3} shows the PHENIX detector at RHIC.
Both central arms (East and West) of the detector include Drift Chamber (DC),
Pad Chamber (PC), Ring Imaging \v{C}erenkov Counter (RICH), lead-scintillator
sandwich type (PbSc) and lead-glass type (PbGl) calorimeter. The East arm has
a Time Of Flight (TOF) detector and a Time Expansion Chamber (TEC) in addition.
Each covers a pseudo-rapidity range of $|\eta|<$0.35 and a quarter azimuth.
A detailed description of the detector is given in the literature~\cite{bib9}.

\begin{figure}[htbp]
\begin{minipage}{44mm}
\leavevmode\epsfxsize=4.4cm
\epsfbox{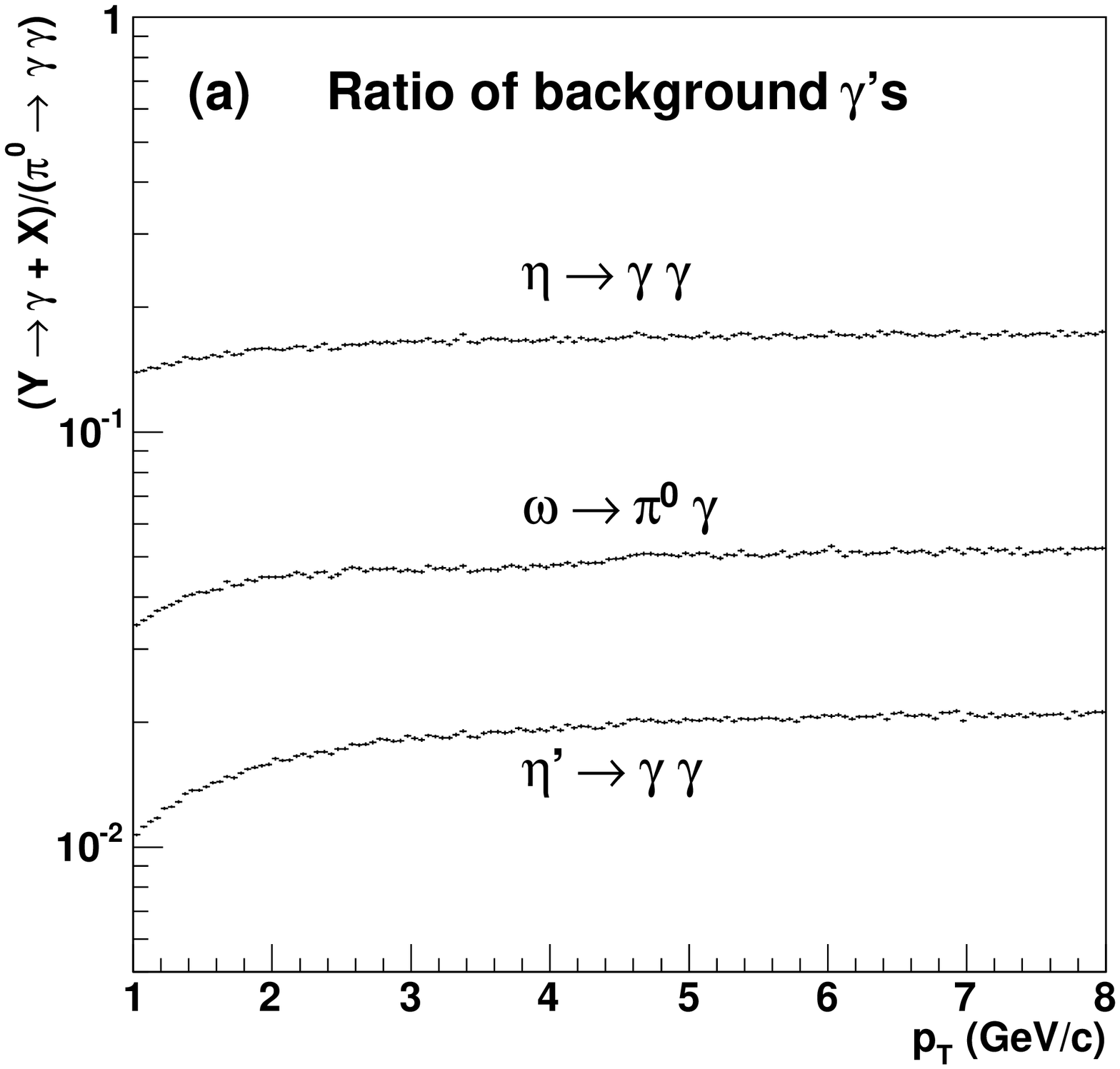}
\end{minipage}
\hspace{5mm}
\begin{minipage}{85mm}
\leavevmode\epsfxsize=8.2cm
\epsfbox{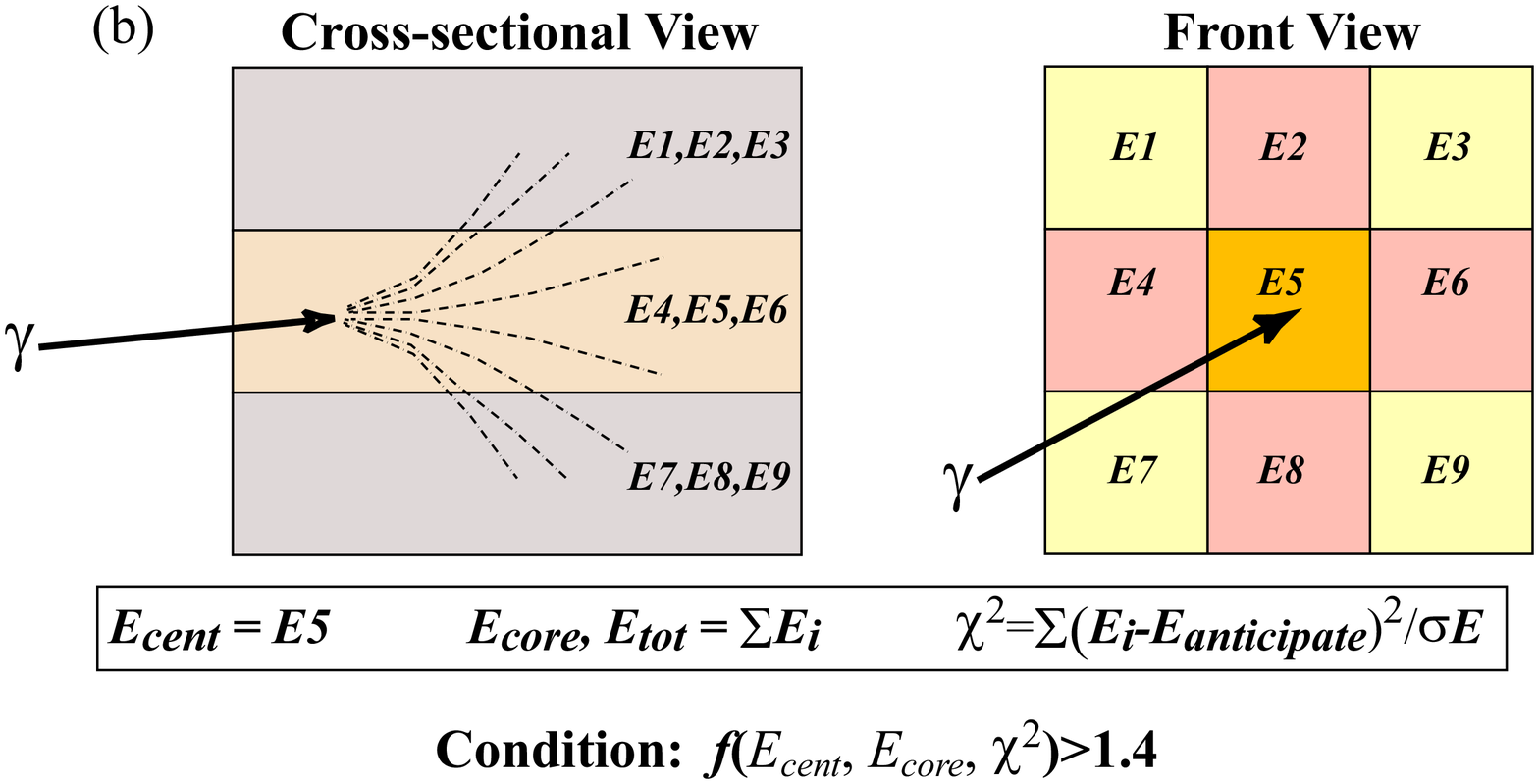}
\end{minipage}
\caption[]{(a) Ratios of $\gamma$'s from hadrons to $\gamma$'s from $\pi^0$, and (b) Quantities used for photon identification.}
\label{fig4}
\end{figure}

In this analysis, PbSc was used for measuring energies of photons and
$\pi^0$'s. The analysis of direct photons require a precise determination
of background photons decaying from known hadronic sources such as
$\pi^0$ and $\eta$. PHENIX has measured the transverse momentum spectra of
$\pi^0$ up to 20\,GeV/$c$ in Au+Au collisions at $\sqrt{s_{NN}}$=200\,GeV
with $\sim$900\,M minimum bias events from RHIC 2004 run~\cite{bib10}.
This work used $\sim$93\,M events out of the above in order to control
run-by-run systematic fluctuation. The momentum spectra of $\eta$ and other
hadronic sources were estimated by replacing $p_T$ by
$({p_T}^2-{M_{\pi}}^2+{M_{h}}^2)^{1/2}$ in fit functions to $\pi^0$ spectra.
Normalization factors for the estimated function were determined so that
${\eta/\pi^0}_{p_T= \infty}$=0.45$\pm$0.05~\cite{bib5},
${\eta'/\pi^0}_{p_T= \infty}$=1.0 and ${\omega'/\pi^0}_{p_T= \infty}$=1.0.
The ratio of background photons are shown in Fig.~\ref{fig4}(a).

Photon candidates were selected by applying a threshold on a PID likelihood
variable computed from several quantities such as cluster shower shape or ratio
of energies among towers (Fig.~\ref{fig4}(b)). It resulted in a significant
reduction of hadronic clusters in the samples. The remaining hadron
contamination and PID efficiency were corrected for based on a simulation study.
Since the ratio of $\gamma$ to $\pi^0$ cancels their common systematic errors,
the excess of the measured photon over the estimated background photon
is evaluated in terms of
$(\gamma/\pi^0)_{measured}/(\gamma/\pi^0)_{background}$ (double ratio).
The systematic errors are summarized in Tab.~\ref{tab1}. The total
$p_T$-correlated error is 7.5\,\%, and the total point-by-point systematic
error is 7.0\,\%, respectively.
\begin{table}[htbp]
\caption{Systematic Errors on direct photon and very low mass dilepton measurement}
\label{tab1}
\begin{minipage}{65mm}
\begin{tabular}{c|c}
\multicolumn{2}{c}{Direct photon measurement} \\
\hline\hline
$\pi^0$ peak extraction & 6\,\% \\
Energy scale in terms of $\gamma/\pi^0$ & 3\,\% \\
$\gamma/\pi^0$ background &  3\,\% \\
$\gamma$ unfolding & 3\,\% \\
Acceptance in terms of $\gamma/\pi^0$ & 2\,\% \\
PID error in terms of $\gamma/\pi^0$ & 1.5\,\% \\
Off-vertex contribution to $\gamma/\pi^0$ & 2.1\,\% \\
Hadron contamination estimate & 5\,\% \\ 
$\gamma/\pi^0$ conversion correction & 1.5\,\% \\
$\eta/\pi^0$ estimate error & 2\,\% \\ \hline \hline
\end{tabular}
\end{minipage}
\begin{minipage}{65mm}
\begin{tabular}{c|c}
\multicolumn{2}{c}{}\\
\multicolumn{2}{c}{}\\
\hline \hline
Total $p_T$-correlated error & 7.5\,\% \\
Total point-by-point error & 7.0\,\% \\ \hline \hline
\multicolumn{2}{c}{}\\
\multicolumn{2}{c}{Very low mass dilepton measurement}\\
\multicolumn{2}{c}{Numbers are relative error on $\gamma^{*}_{direct}/\gamma^{*}_{all}$}\\
\hline\hline
$h/\pi^0$ estimate & 20\,\% \\
Acceptance & 5\,\% \\
Total error & 21\,\% \\ 
(point-by-point only) & \\
\hline\hline
\multicolumn{2}{c}{}\\
\end{tabular}
\end{minipage}
\end{table}

\begin{figure}[htbp]
\centering
\leavevmode\epsfxsize=7.0cm
\epsfbox{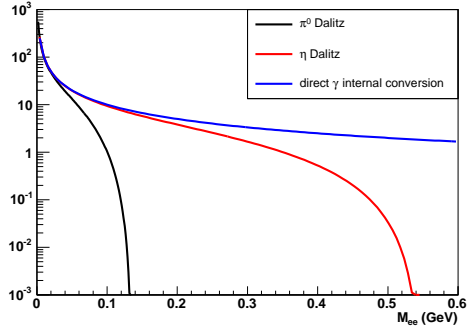}
\caption{Invariant mass spectra by the Dalitz decay of $\pi^0$, $\eta$ and internal conversion of direct photons.}
\label{fig5_1}
\end{figure}

In addition to the above conventional approach, a new method that utilizes
very low mass dileptons has been applied~\cite{bib11}. It is well known
that the $e^+e^-$ invariant mass distributions from $\pi^0$ or $\eta$
Dalitz decay follow the Kroll-Wada formula as shown below~\cite{bib12}.
\[\frac{1}{N_{\gamma}} \frac{dN_{ee}}{dm_{ee}} = \frac{2 \alpha}{3 \pi} \sqrt{1-\frac{4 m_{e}^2}{m_{ee}^2}}(1+\frac{2m^2_e}{m^2_{ee}})\frac{1}{m_{ee}}|F(m^2_{ee})|^2 (1-\frac{m^2_{ee}}{M^2})^3 \]
where $m_{e}$ is the mass of electrons, $m_{ee}$ is the invariant mass of a
produced dilepton, $M$ and $|F(m^2_{ee})|$ are the mass and the form factor of
a parent particle, respectively. The formula is also valid for internal
conversion of direct photons by substituting $|F(m^2_{ee})|$ with 1. The
$e^+e^-$ invariant mass distribution of $\pi^0$ and $\eta$ Dalitz decays as
well as direct photon internal conversion are demonstrated in Fig.~\ref{fig5_1}.
Using the formula, the amount of dileptons
can be estimated at a given invariant mass region where a ratio
of the measured to the background is taken as a function of $p_T$ (in this
analysis, 90$<$$M_{ee} $$<$300\,MeV/$c^2$).
Since the yield at the lowest mass region is approximately proportional to
the total number of produced $e^+e^-$, the ratio at the measured mass region
is then converted into the one at the very low mass region
(0$<$$M_{ee}$$<$30\,MeV/$c^2$) using the Kroll-Wada formula.
This work used $\sim$900\,M minimum bias events.
The systematic errors for this method are also summarized in Tab.~\ref{tab1}.

\section{Results and discussion}
\begin{figure}[htbp]
\begin{minipage}{44mm}
\leavevmode\epsfxsize=4.4cm
\epsfbox{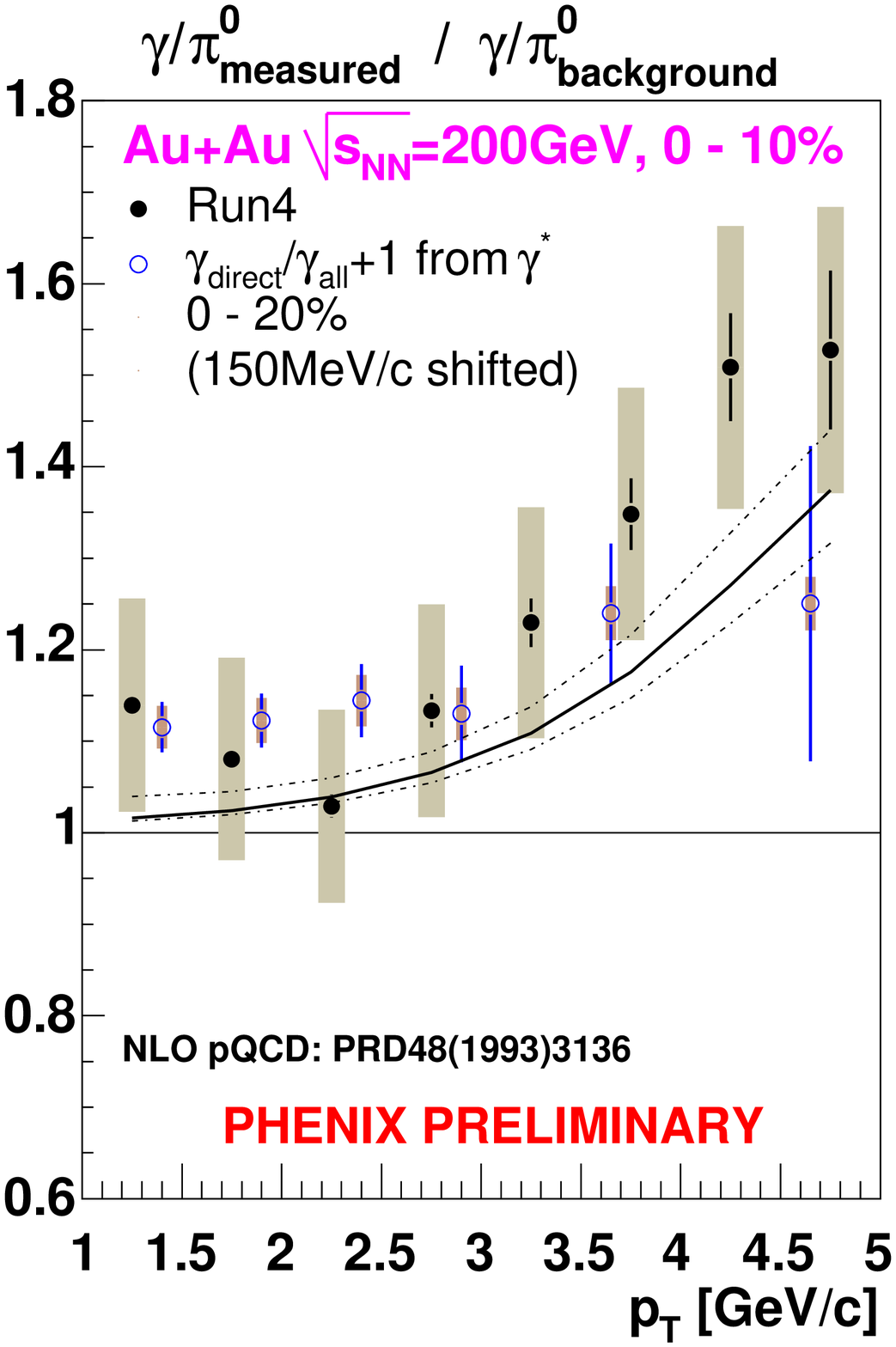}
\end{minipage}
\begin{minipage}{44mm}
\leavevmode\epsfxsize=4.4cm
\epsfbox{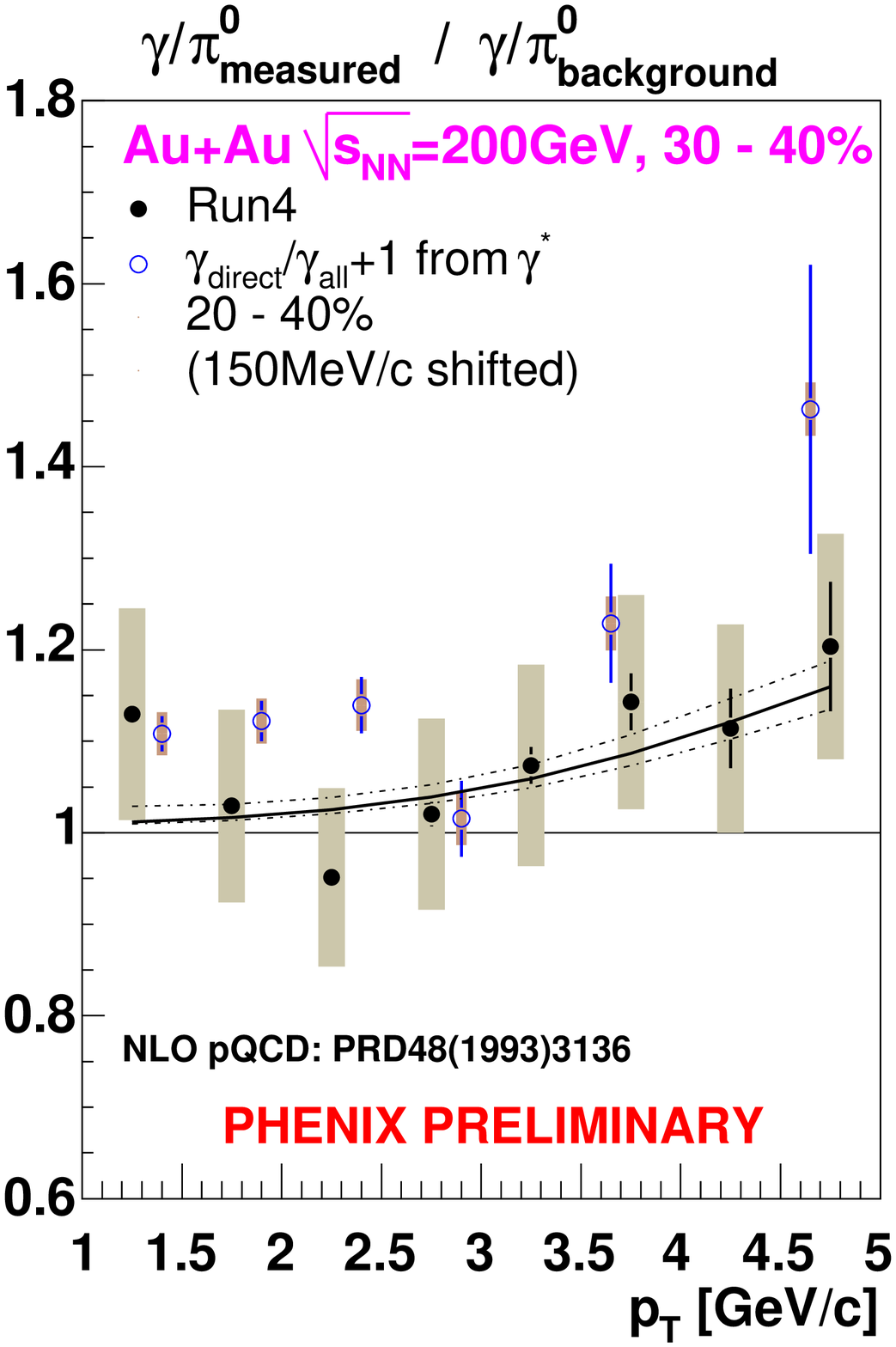}
\end{minipage}
\begin{minipage}{44mm}
\leavevmode\epsfxsize=4.4cm
\epsfbox{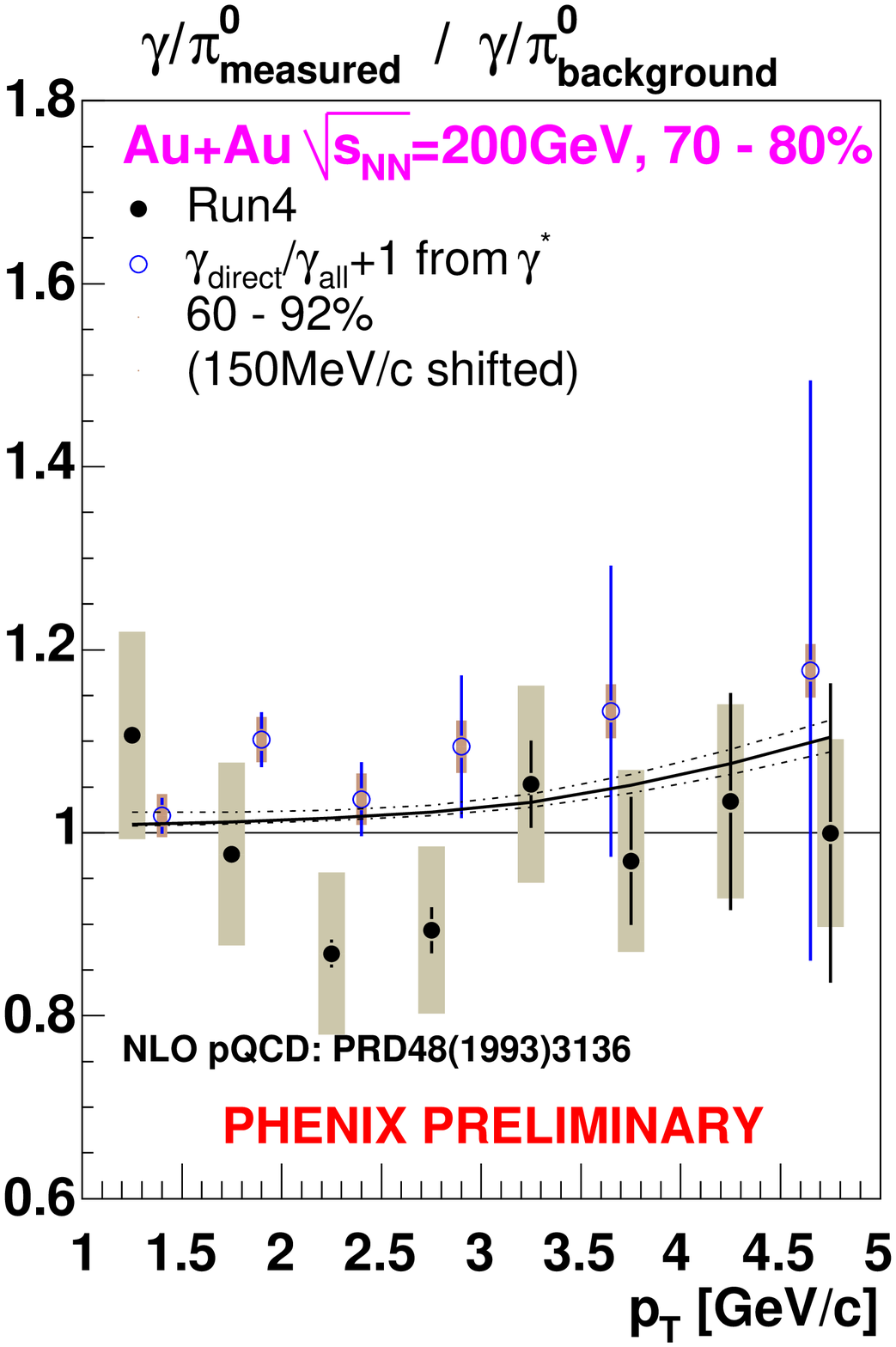}
\end{minipage}
\vspace{-3mm}
\caption[]{$(\gamma/\pi^0_{measured})/(\gamma/\pi^0_{background})$ for real photon measurement for 0-10\,\% (left), 30-60\,\% (middle) and 70-80\,\% (right) centrality (black solid points), together with $\gamma^*_{direct}/\gamma^*_{all} +1$ from very low-mass dilepton analysis (blue open points), and NLO pQCD expectation scaled by number of binary nucleon-nucleon collisions (lines).}
\label{fig6}
\end{figure}

Fig.~\ref{fig6} shows the $\gamma/\pi^0$ double ratio for 0-10\,\% central
collisions as a function of $p_T$.
The results from very low mass dilepton analysis are also plotted on the
figures in blue, in the form of $\gamma^*_{direct}/\gamma^*_{all}+1$, which is
not equivalent to $\gamma/\pi^0$ double ratio. Both direct photons and
very low mass dileptons are consistent given the large error of the direct
photon measurement. They are also consistent with the published results from
RHIC 2002 run~\cite{bib5}. In 1$<$$p_T$$<$2.5\,GeV/c, there is no significant
excess in the direct photon measurement in all the centralities, while
the very low mass dilepton has a significant excess except for peripheral
events. Overlaid on the plots are expectations from a NLO pQCD calculation
scaled by the number of binary nucleon-nucleon collisions~\cite{bib13}.
The direct photon results can be well described by the expectations except
the 0-10\,\% centrality data, where the onset of an additional contribution
is seen in 4$<$$p_T$$<$5\,GeV/c.

\begin{figure}[htbp]
\begin{center}
\leavevmode\epsfxsize=4.4cm
\epsfbox{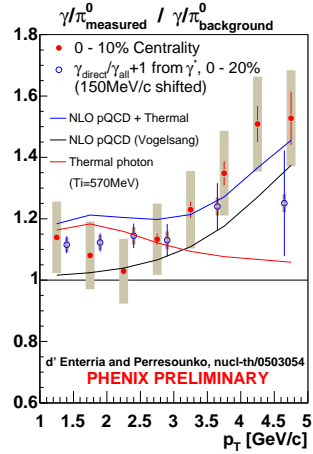}
\vspace{-3mm}
\caption[]{0-10\,\% central collisions compared with a model including thermal radiation from QGP.}
\label{fig7}
\end{center}
\end{figure}

The result for most central events is also compared with a model including
the thermal radiation, and shown in Fig.~\ref{fig7}.
The model assumes an initial temperature of $T_0^{ave}$ = 360\,MeV
($T_0^{max}$ = 570\,MeV) and a formation time of
$\tau_0$= 0.15\,fm/$c$~\cite{bib14}. The result is not inconsistent with
the model, suggesting that there is a possibility of the existence of
a thermal source.
However, further works on reducing systematic errors on direct photon
measurement, and confirmation of no excess in very low mass dilepton
in p+p measurement are desired to make a concrete statement.

\section{Conclusion}
Direct photon production in Au+Au collisions at $\sqrt{s_{NN}}$=200\,GeV
has been measured in the $p_T$ range of 1$<$$p_T$$<$5\,GeV/$c$. The yield
as a function of centrality is in agreement with published data of the
RHIC 2002 run, and with results from a new method that explores very low
mass dileptons. The result is compared to several theoretical calculations,
and it is found that the measurement is not inconsistent with calculations
including thermal photon contributions.
Further works on reducing systematic errors on direct
photon measurement, and confirmation of no excess in very low mass dilepton
in p+p measurement are desired to make a concrete statement.


\begin{thebibliography}{99}
\bibitem{bib1} I. Arsene, et al. (BRAHMS Coll.), {\it Nucl. Phys.} {\bf A757} (2005) 1.
\bibitem{bib2} B.B. Back, et al., (PHOBOS Coll.), {\it Nucl. Phys.} {\bf A757} (2005) 28.
\bibitem{bib3} J. Adams, et al. (STAR Coll.), {\it Nucl. Phys.} {\bf A757} (2005) 102.
\bibitem{bib4} K. Adcox, et al. (PHENIX Coll.), {\it Nucl. Phys.} {\bf A757} (2005) 184.

\bibitem{bib5} S.S. Adler, et al. (PHENIX Coll.), {\it Phys. Rev. Lett.} {\bf 94}, 232301 (2005).

\bibitem{bib6} S.S. Adler, et al. (PHENIX Coll.), accepted by {\it Phys. Rev. Lett.}, nucl-ex/0510047.

\bibitem{bib7} R.J. Fries, B. M\"{u}ller and  D.K. Srivastava, {\it Phys. Rev.} {\bf C72}, 041902 (2005)

\bibitem{bib8} S. Turbide, R. Rapp and C. Gale, {\it Phys. Rev.} {\bf C69}, 140903 (2004).

\bibitem{bib9} K. Adcox, et al., (PHENIX Coll.), {\it Nucl. Inst. \& Meth. Phys. Res.} {\bf A499} (2003) 469.
\bibitem{bib10} T. Isobe (PHENIX Coll.), nucl-ex/0510085.

\bibitem{bib11} S. Bathe (PHENIX Coll.), nucl-ex/0511042.

\bibitem{bib12} N.M. Kroll and W. Wada, {\it Phys. Rev.}, {\bf 98}, 1355 (1955).

\bibitem{bib13} L.E. Gordon and W. Vogelsang, {\it Phys. Rev.} {\bf D48}, 3136 (1993).

\bibitem{bib14} D. d'Enterria and D. Peressounko, nucl-th/0503054.

\end{thebibliography}
\end{document}